# FLUID-ELASTIC COEFFICIENTS IN SINGLE PHASE CROSS FLOW: DIMENSIONAL ANALYSIS, DIRECT AND INDIRECT EXPERIMENTAL METHODS

**Romain Lagrange**
DEN-Service d'Etudes Mécaniques et Thermiques
SEMT, CEA, Université Paris Saclay
F-91191 Gif-Sur-Yvette, France

**Philippe Piteau**
DEN-Service d'Etudes Mécaniques et Thermiques
SEMT, CEA, Université Paris Saclay
F-91191 Gif-Sur-Yvette, France

**Xavier Delaune**
DEN-Service d'Etudes Mécaniques et Thermiques
SEMT, CEA, Université Paris Saclay
F-91191 Gif-Sur-Yvette, France

**Jose Antunes**
Centro de Ciências e Tecnologias Nucleares
Instituto Superior Técnico, Universidade de Lisboa
Estrada Nacional 10, Km 139.7
2695-066 Bobadela LRS, Portugal

## ABSTRACT

The importance of fluid-elastic forces in tube bundle vibrations can hardly be over-emphasized, in view of their damaging potential. In the last decades, advanced models for representing fluid-elastic coupling have therefore been developed by the community of the domain. Those models are nowadays embedded in the methodologies that are used on a regular basis by both steam generators providers and operators, in order to prevent the risk of a tube failure with adequate safety margins. From an R&D point of view however, the need still remains for more advanced models of fluid-elastic coupling, in order to fully decipher the physics underlying the observed phenomena. As a consequence, new experimental flow-coupling coefficients are also required to specifically feed and validate those more sophisticated models. Recent experiments performed at CEA-Saclay suggest that the fluid stiffness and damping coefficients depend on further dimensionless parameters beyond the reduced velocity.

In this work, the problem of data reduction is first revisited, in the light of dimensional analysis. For single-phase flows, it is underlined that the flow-coupling coefficients depend at least on two dimensionless parameters, namely the Reynolds number $Re$ and the Stokes number $Sk$. Therefore, reducing the experimental data in terms of the compound dimensionless quantity $V_r = Re/Sk$ necessarily leads to impoverish results, hence the data dispersion. In a second step, experimental data are presented using the dimensionless numbers $Re$ and $Sk$. We report experiments, for a 3x5 square tube bundle subjected to water transverse flow. The bundle is rigid, except for the central tube which is mounted on a flexible suspension allowing for translation motions in the lift direction.

The evolutions of the flow-coupling coefficients with the flow velocity are determined using two different experimental procedures: (1) In the direct method, an harmonic motion of increasing frequency is imposed to the tube. (2) In the indirect method, the coefficients are obtained from the modal response of the tube (frequency, damping). The coefficient identification was performed well beyond the system instability boundary, by using active control, allowing an exploration of a significant range of flow velocity.

For a given $Sk$, the results show that: (a) at low $Re$, the flow-coupling coefficients are close to zero; (b) at intermediate $Re$, the flow stabilizes the tube; (c) at high $Re$, the flow destabilizes the tube, leading to a damping-controlled instability at a critical $Re$. Reducing the data in terms of $Re$ and $Sk$ clarifies the various experimental "branches", which are mixed when using $V_r$. The two identification techniques lead to reasonably compatible fluid-elastic coefficients.

## NOMENCLATURE

$$c_D = \frac{-2\widehat{C_f}}{\rho_f DLV_g}$$ Dimensionless damping coefficient

$$c_K = \frac{-2K_f}{\rho_f LV_g^{\,2}}$$ Dimensionless stiffness coefficient

$$c_T = \frac{C_s + C_f}{C_s + C_f^0}$$ Dimensionless total damping coefficient

$C_f$ Fluid-added damping

$C_f^0$ Fluid-added damping in still fluid

$\widehat{C_f} = C_f - C_f^0$     Fluid-added damping due to fluid velocity

$C_s$     Structural damping of the moving tube

$D$     Tube diameter

$F_a$     Excitation fluid force

$F_0$     Frequency of first mode in still fluid

$K_f$     Fluid-added stiffness

$K_s$     Structural stiffness of the moving tube

$L$     Tube length

$l = L/D$     Tube aspect ratio

$M_s$     Structural mass of the moving tube

$M_f$     Fluid-added mass

$P$     Tube bundle pitch

$p = P/D$     Pitch ratio

$Re = D V_g / \nu$     Reynolds number

$Sk = D^2 F_0 / \nu$     Stokes number

$Sc = 2\pi\xi_0 \dfrac{M_s + M_f}{\rho_f D^2 L}$     Scruton number

$V_g$     Gap flow velocity

$V_r = V_g / (D F_0)$     Reduced velocity

$X$     Tube modal displacement

$\rho_f$     Mass density of the fluid

$\nu$     Kinematic viscosity of the fluid

$\xi_0$     Damping of first mode in still fluid

$\lambda$     Characteristic length

$M$     Characteristic mass

$\tau$     Characteristic time

## INTRODUCTION

The knowledge of the fluid force acting on a structure subject to a cross flow is a crucial information that must be accounted for when designing heat-exchanger tube bundles. The large vibrations resulting from a fluid-elastic instability may lead to some mechanical degradation of the concerned tube, which may affect the power plant operation and safety. This instability can be described as a self-excited feedback mechanism between the motion of the structure and the fluid forces. Since the pioneering work of Tanaka and Takahara [1], several authors [2-10] measured the fluid-elastic force to feed the stability criterion models developed by Connors [11], Blevins [12], Chen [2,3], Lever and Weaver [13-15], Price and Païdoussis [16-18], Granger et al. [19,20] and Tanaka et al. [21].

Still, further experimental work is needed to accurately understand the effects of changing the tube mass, damping, frequency or diameter, as well as the bundle configuration, even for a single flexible tube within a rigid bundle subject to single-phase flow.

There are basically two techniques for obtaining the fluid-elastic force. The direct method [1,7-10] is based on imposing a controlled oscillatory motion to a given tube within a rigid bundle and measuring the forces exerted by the flow as a function of the flow velocity and of the motion frequency. The fluid-elastic force is obtained from the transfer function between the tube displacement and the measured force from which the structural inertia term is subtracted. The indirect method [4-6,22,23] applies to an instrumented tube, for which the fluid-elastic force is extracted from the changes in the modal frequency and damping of the vibrating tube. Both techniques have advantages and drawbacks, concerning the complexity of the setup, the sensitivity to external perturbations and the range of parameters that may be explored. Given all these constraints, an experimental rig allowing both measuring methods, as well as exploring beyond the instability threshold, has been developed at CEA, see [4-6, 22,23]. In this paper, we analyse these experimental results in the light of a dimensional analysis. In particular, we investigate the influence of the tube frequency and the water flow velocity on the fluid-elastic force.

## DIMENSIONAL FLUID-ELASTIC COEFFICIENTS

We consider the vibration of a flexibly mounted tube (diameter $D$, length $L$), part of a 3x5 square tube bundle (pitch $P$), immersed in a viscous fluid of volume mass density $\rho_f$, kinematic viscosity $\nu$ and gap velocity $V_g$, see Fig. 1. We note $M_s$, $C_s$ and $K_s$ the mass, damping and rigidity coefficients of the first mode of vibration of the flexible tube in air. The modal displacement $X$ in the lift direction is assumed to satisfy the equation:

$$(M_s + M_f)\ddot{X} + (C_s + C_f)\dot{X} + (K_s + K_f)X = F_a, \qquad (1)$$

where $F_a$ is an excitation fluid force considered as independent on the tube motion. The modal frequency of the flexible tube in still water is

$$F_0 = \sqrt{K_s / (M_s + M_f)} / (2\pi) \cdot \qquad (2)$$

As most often assumed, we postulate that the added mass $M_f$ does not depend on flow velocity, but only on fluid density and bundle geometry. In other words, $M_f$ is an unknown function $H_{M_f}$ of $(\rho_f, D, L, P)$ :

$$M_f = H_{M_f}\left(\rho_f, D, L, P\right).  \qquad (3)$$

On the other hand, we consider that the fluid added damping $C_f$ and rigidity $K_f$ depend, at least, on the flexible tube frequency $F_0$, the fluid material properties $\rho_f, \nu$, the gap velocity $V_g$, as well as the bundle geometry

$$C_f = H_{C_f}\left(F_0, \rho_f, \nu, V_g, D, L, P\right),  \qquad (4)$$

$$K_f = H_{K_f}\left(F_0, \rho_f, \nu, V_g, D, L, P\right).  \qquad (5)$$

Let $C_f^0$ be the fluid added damping coefficient in still fluid, defined as the value of $C_f$ as $V_g = 0$. Then, $\widehat{C_f} = C_f - C_f^0$ is a measure of the effect of the fluid velocity $V_g$ on the fluid added damping coefficient $C_f$, and is a function of $\left(F_0, \rho_f, \nu, V_g, D, L, P\right)$:

$$\widehat{C_f} = H_{\widehat{C_f}}\left(F_0, \rho_f, \nu, V_g, D, L, P\right).  \qquad (6)$$

The relations of dependence (3) to (6) constitute a minimal model, based on experimental observations, bibliography reporting and physical intuition. More advanced models would also consider the effect of some other parameters, for e.g. the roughness of the tubes.

## DIMENSIONLESS FLUID-ELASTIC COEFFICIENTS

The dimensional analysis is based on the Vaschy-Buckingham theorem. The theorem states that an equation involving $n$ physical variables with $k$ fundamental units (usually $k = 3$ in classical mechanics) can be reduced to an equation involving $n - k$ dimensionless parameters. Thus, introducing a scale of length $\lambda$, mass $M$ and time $\tau$, the equations (3), (5) and (6) are physically meaningful if

$$\frac{M_f}{M} = H_{M_f}\left(\frac{\rho_f}{M\lambda^{-3}}, \frac{D}{\lambda}, \frac{L}{\lambda}, \frac{P}{\lambda}\right),  \qquad (7)$$

$$\frac{K_f}{M\tau^{-2}} = H_{K_f}\left(\frac{F_0}{\tau^{-1}}, \frac{\rho_f}{M\lambda^{-3}}, \frac{\nu}{\lambda^2\tau^{-1}}, \frac{V_g}{\lambda\tau^{-1}}, \frac{D}{\lambda}, \frac{L}{\lambda}, \frac{P}{\lambda}\right),  \qquad (8)$$

$$\frac{\widehat{C_f}}{M\tau^{-1}} = H_{\widehat{C_f}}\left(\frac{F_0}{\tau^{-1}}, \frac{\rho_f}{M\lambda^{-3}}, \frac{\nu}{\lambda^2\tau^{-1}}, \frac{V_g}{\lambda\tau^{-1}}, \frac{D}{\lambda}, \frac{L}{\lambda}, \frac{P}{\lambda}\right).  \qquad (9)$$

As (7) involves five dimensional quantities with two fundamental dimensions (length and mass), it can be reduced to a relation between three dimensionless quantities. Similarly, as (8) and (9) involve eight dimensional quantities with three fundamental dimensions (length, mass and time), they can be reduced to a relation between five dimensionless quantities. These dimensionless quantities are not unique and derive from a specific choice for the characteristic length $\lambda$, mass $M$ and time $\tau$. Picking $\lambda = D$, $M = \rho_f D^2 L$ and $\tau = D/V_g$, the dimensionless equations rewrite

$$\frac{M_f}{\rho_f D^2 L} = H_{m_f}\left(l, p\right),  \qquad (10)$$

$$c_K = \frac{-2K_f}{\rho_f L V_g{}^2} = H_{k_f}\left(V_r, l, Re, p\right),  \qquad (11)$$

$$c_D = \frac{-2\widehat{C_f}}{\rho_f D L V_g} = H_{c_f}\left(V_r, l, Re, p\right),  \qquad (12)$$

with $l = L/D$, $p = P/D$, $Re = D V_g/\nu$, and $V_r = V_g/D F_0$ the tube aspect ratio, the pitch ratio, the Reynolds number and the reduced velocity, respectively. In what follows, we shall also make use of the Stokes number, obtained from the ratio between $Re$ and $V_r$: $Sk = Re/V_r = D^2 F_0/\nu$. To study the stability of the tube, we introduce the total damping coefficient $c_T = \left(C_s + C_f\right)/\left(C_s + C_f^0\right)$, which also rewrites as

$$c_T = \frac{C_s + C_f^0 + \widehat{C_f}}{C_s + C_f^0} = 1 + \frac{\widehat{C_f}}{C_s + C_f^0} = 1 + \frac{c_D \rho_f D L V_g}{-2\left(C_s + C_f^0\right)}.  \qquad (13)$$

Introducing $Sc = 2\pi\xi_0\left(M_s + M_f\right)/\left(\rho_f D^2 L\right)$ and $\xi_0 = \left(C_s + C_f^0\right)/\left[2\left(M_s + M_f\right)\left(2\pi F_0\right)\right]$ as the Scruton number and the reduced damping parameter for a flexible tube in a still fluid, (13) simplifies to

$$c_T = 1 - \frac{c_D V_r}{4 Sc} = H_{c_T}\left(V_r, l, Re, p, Sc\right).  \qquad (14)$$

It follows that the tube is stable if $c_T > 0$, unstable if $c_T < 0$ and the stability threshold $c_T = 0$ is a function of $\left(V_r, l, Re, p, Sc\right)$.

The above dimensionless analysis shows that the fluid-elastic coefficients $c_D$ and $c_K$ are not a function of $V_r$ only, but also depend on the Reynolds number. The reduced velocity has the disadvantage to encapsulate both $F_0$ and $V_g$ in a same dimensionless number, whereas these two parameters are independent. Consequently, to distinguish the effect of the frequency and the fluid flow on the variation of the fluid-elastic coefficient, it is preferable to use the couple of dimensionless numbers $(Sk, Re)$ instead of $(V_r, Re)$. Adopting this point of view in the next sections, $c_D$ and $c_K$ are seen as functions of $(Sk, l, Re, p)$. Similarly, the total damping coefficient $c_T$ depends on $(Sc, Sk, l, Re, p)$.

## EXPERIMENTAL SETUP AND MEASUREMENT METHODS

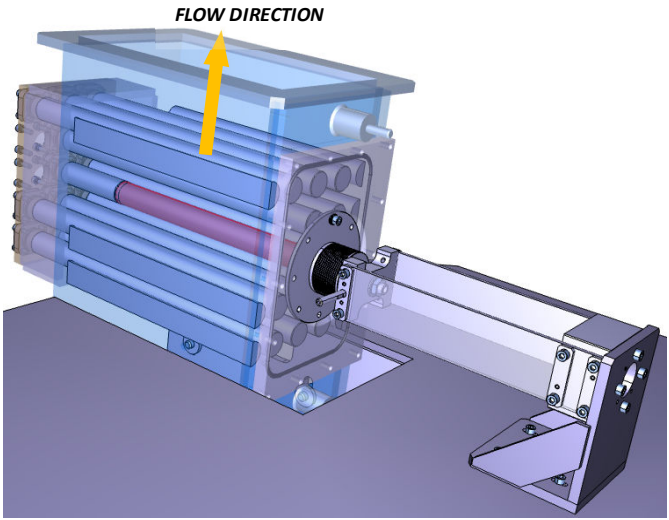

**Figure 1. Bundle and flexibly mounted tube.**

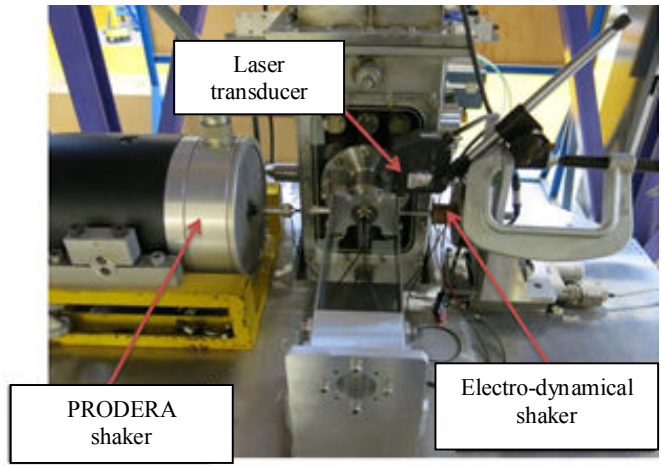

**Figure 2. Experimental setup.**

An experimental setup, has been built at CEA to study the variations of the fluid-elastic coefficients $c_D$ and $c_K$. This experimental setup is sketched in figures 1 and 2 and described briefly in the following. Readers should refer to [4-6,22,23] for an extensive description. The tube bundle has immersed length $L = 300$ mm and diameter $D = 30$ mm, with a pitch ratio $P/D = 1.5$. As depicted in Fig. 1, the moving tube is in the central position of a $3 \times 5$ square bundle made of rigid tubes (plus two columns of 5 half-tubes at the boundaries). The central tube is supported by two parallel flexible steel blades allowing large vibrations in the lift direction. The tube bundle is subject to a cross water flow with a gap speed in the range $V_g = 0 \sim 6$ m/s. In terms of dimensionless numbers, this range of variation for $V_g$ yields $Re = 0 \sim 10^5$. The fluid-elastic coefficients $c_D$ and $c_K$ are measured using two experimental approaches.

In the indirect method, the motion of the flexible tube is free and the measurement of its displacement and velocity is provided by a laser transducer (Keyence LK-G500). In order to investigate the variations of the fluid-elastic coefficients well beyond the stability threshold, a feedback control loop made of an electro-dynamical shaker was used, see [4-6,23]. The coefficients $c_D$ and $c_K$ are obtained from the variations of the tube motion frequency and damping. This technique for extraction of the fluid-elastic coefficients from the modal parameters is simple and proved robust enough. Since three coefficients are to be extracted from only two modal parameters, the assumption of a constant added mass (e.g. a velocity-independent flow inertia coefficient) has to be somewhat arbitrarily enforced, which is a disadvantage of this identification method. Any possible changes in $M_f$ due to the fluid velocity will then be reflected in the coupling coefficient $K_f$. The modal frequency of the moving tube is set by changing the thickness of the blades or by adding a suitable mass to the moving fixture. Several configuration tests, denoted L1, L2, L3, L3B, L4B and L4TI, see Table 1, have been performed, covering a large range of frequencies $F_0 \approx 13 \sim 39$ Hz. The second modal frequency is much larger than $F_0$, such that the tube dynamics is mainly a rigid translation.

In the direct method, an harmonic motion of imposed frequency is directly applied to the tube, thanks to a PRODERA shaker, see Fig. 2. The fluid force acting on the flexible tube is only measured in the lift direction with a KISTLER sensor. The coefficients $c_D$ and $c_K$ are directly extracted from the measure of the fluid force. This method, first introduced by Tanaka in the 80's and mainly followed by Chen in the 90's has yield some interesting results. However, despite its apparent simplicity, it has been progressively abandoned due to its difficult experimental implementation. Three measurement campaigns, denoted L1TI 2017, L1TI 2018 and L3TI have been conducted, with imposed frequencies covering the range $F_0 \approx 13 \sim 39$ Hz .

| TEST | $\dfrac{M_t + M_f}{L}$ (kg.m$^{-1}$) | $\xi_0$ (%) | $F_0$ (Hz) | Scruton $Sc$ ($\times 10^{-1}$) | Stokes $Sk$ ($\times 10^4$) |
|------|------|------|------|------|------|
| L1 | 3.77 | 0.90 | 12.99 | 2.36 | 1.169 |
| L2 | 4.57 | 1.19 | 26.47 | 3.78 | 2.382 |
| L3 | 5.17 | 0.71 | 38.58 | 2.57 | 3.472 |
| L3B | 6.77 | 0.69 | 33.72 | 3.27 | 3.035 |
| L4 | 4.17 | 0.64 | 20.76 | 1.86 | 1.868 |
| L4B | 6.10 | 0.62 | 17.47 | 2.64 | 1.572 |
| L4TI | 5.33 | 0.61 | 18.40 | 2.27 | 1.656 |

**Table 1. Tested configurations. Indirect method.**

## EXPERIMENTAL RESULTS

In this section, we present the experimental measurements of $c_D$, $c_K$ and $c_T$, obtained from the indirect and direct methods. We focus on the evolution of these dimensionless coefficients with the Reynolds and the Stokes numbers. On the following figures, curves with identical colours correspond to configuration tests with similar Stokes numbers, see Table 1.

The measurements of $c_D$ obtained from the indirect method are shown in Fig. 3. Whatever the Stokes number, a clear general trend is observed. At low Reynolds numbers, $c_D \approx 0$, such that the fluid velocity has a negligible effect on the stability of the tube. At intermediate Reynolds numbers, $c_D$ becomes negative such that the tube is getting stabilized. In this range of $Re$, some kinetic energy is conveyed by the vibrating tube to the fluid, which in turn propagates this energy through the far domain. This corresponds to an energy loss for the tube leading to a damping of its vibrations. It is believed that the energy propagation through the far domain is enhanced by some fluid vortices whose existence still needs to be proven. At high Reynolds numbers, $c_D$ increases and the tube becomes unstable for some critical $Re_c$, corresponding to $c_T = 0$. In this range of $Re$, some elastic and kinetic energy is conveyed by the fluid motion to the tube. This corresponds to an energy gain for the tube whose vibrations are amplified. In this range, we note that $c_D$ decreases with the Stokes number $Sk$, such that a tube with a high frequency is more stable than a tube with a low frequency.

In figures 4, 5 and 6, we compare the measurements of $c_D$ obtained from the indirect method and the direct method. For $Sk \approx 1,169 \times 10^4$ (i.e. $F_0 \approx 13$ Hz , figure 4) and $Sk \approx 1,7 \times 10^4$ (i.e. $F_0 \approx 18$ Hz , figure 5) the two methods yield similar experimental results. For $Sk \approx 2,382 \times 10^4$ (i.e. $F_0 \approx 26$ Hz , figure 6), significant differences are observed, especially at low Reynolds numbers. We attribute these differences to a bad signal to noise ratio due to some parasitic frequencies in the experimental setup as the forcing frequency $F_0$ of the direct method is increased. Also, at high forcing frequencies, the precise determination of the fluid-elastic force is complicated as most of the measured force has an inertia origin.

The measurements of $c_K$ obtained from the indirect method are shown in Fig. 7. At low Reynolds numbers, the data are scattered and difficult to analyze in the sense that no special trend is clearly observed. However, if the fluid velocity has a negligible effect on the dynamics of the structure, as it is believed from the analysis of $c_D$, then one would expect that $c_K \approx 0$. At intermediate Reynolds numbers, $c_K$ becomes positive such that the relative rigidity of the tube diminishes. As already pointed out, this could be related to the existence of some fluid vortices interacting with the structure. At high Reynolds numbers, $c_K$ decreases such that the relative rigidity of the tube is enhanced. Still, $c_K$ being positive, the total rigidity $K_s + K_f$ of the tube in a flowing fluid is smaller than its rigidity $K_s$ in a fluid at rest. Thus, in this range of $Re$, the frequency of the tube vibrations is smaller than $F_0$.

The evolution of the total damping coefficient $c_T$ is shown in Fig. 8. The variations of this coefficient are directly related to those of $c_D$ through the equation (14). Consequently, at low Reynolds numbers, $c_D \approx 0$ yields $c_T \approx 1$. In this range of $Re$ the fluid velocity does not affect the stability of the structure. At intermediate Reynolds numbers, $c_T$ increases (i.e. $c_D$ decreases), meaning that the tube loses some energy and gets stabilized. At high Reynolds numbers, $c_T$ decreases (i.e. $c_D$ increases) meaning that the tube is gaining some energy from the fluid, leading to amplified vibrations. Eventually, at a critical Reynolds number $Re_c$, the total damping coefficient vanishes and the tube becomes unstable. The figure 8 clearly shows that the stability threshold $Re_c$ increases with the Stokes number $Sk$. The determination of the exact relation between $Re_c$ and $Sk$ is however difficult from the present experimental results as the configuration tests listed in Table 1 do not have exactly the same Scruton numbers. Still, we show in figure 9 that the representation of $c_T$ versus the reduced velocity $V_r = Re/Sk$ does not yield a collapse of the experimental data on a master curve, in particular close to the stability threshold $c_T = 0$. In other words, the flow-coupling coefficients do not only depend on $V_r$, and contrarily to the prediction of Connors [11], $Re_c$ is probably not a linear function of $Sk$.

Finally, the dependence of the fluid-elastic coefficients with the Stokes number might be explained by several physical reasons. The most likely one is that the flow regime might be triggered by the tube frequency. The tube frequency might also affect the time-lag between the tube motion and the fluid forces. Whatever the physical explanation, the results presented suggest a strong dependence on both the flow Reynolds number and the Stokes number, beyond the classicaly assumed dependence on the reduced velocity.

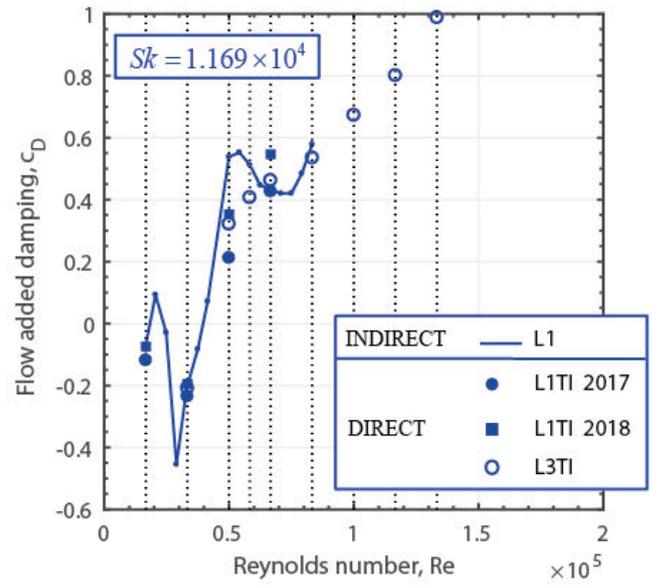

**Figure 4. Evolution of the flow added damping coefficient $c_D$ with the Reynolds number.**

**Indirect and direct methods. $Sk = 1.169 \times 10^4$.**

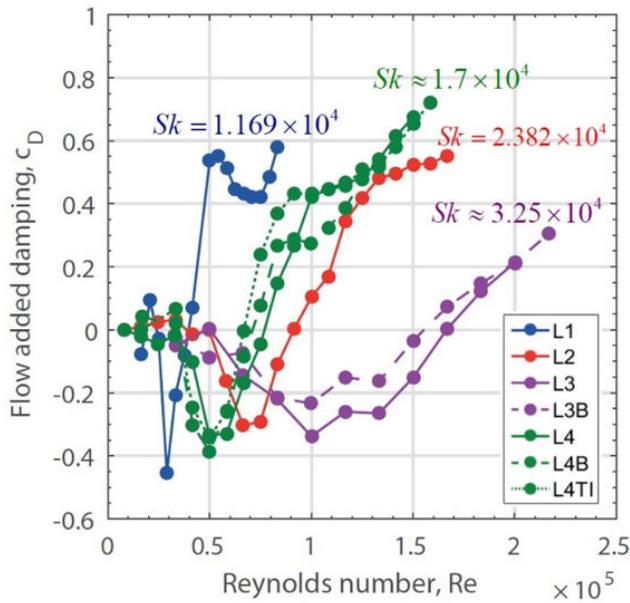

**Figure 3. Evolution of the flow added damping coefficient $c_D$ with the Reynolds and the Stokes numbers. Indirect method.**

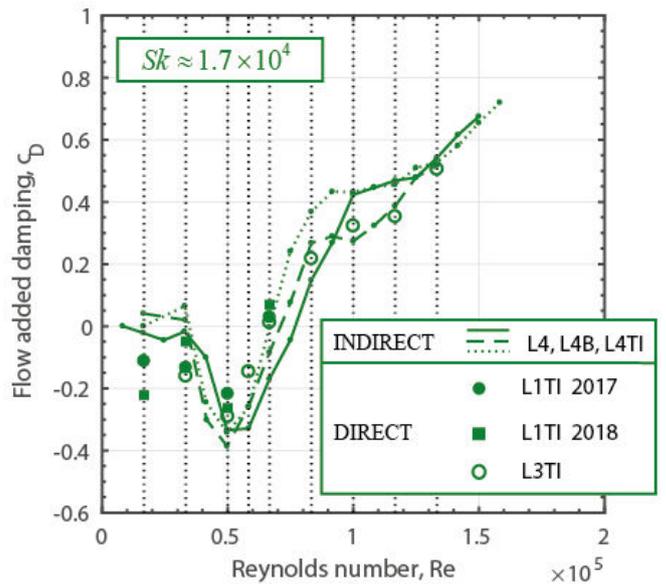

**Figure 5. Evolution of the flow added damping coefficient $c_D$ with the Reynolds number.**

**Indirect and direct methods. $Sk = 1.7 \times 10^4$.**

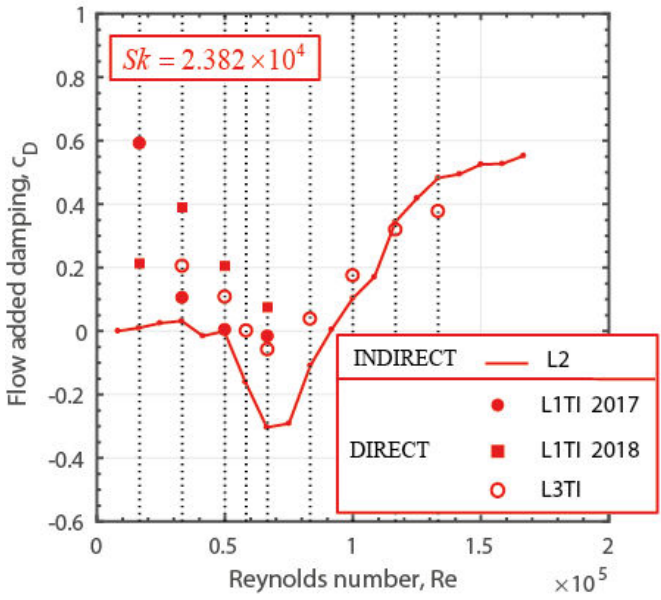

**Figure 6. Evolution of the flow added damping coefficient $c_D$ with the Reynolds number. Indirect and direct methods.** $Sk = 2.382 \times 10^4$.

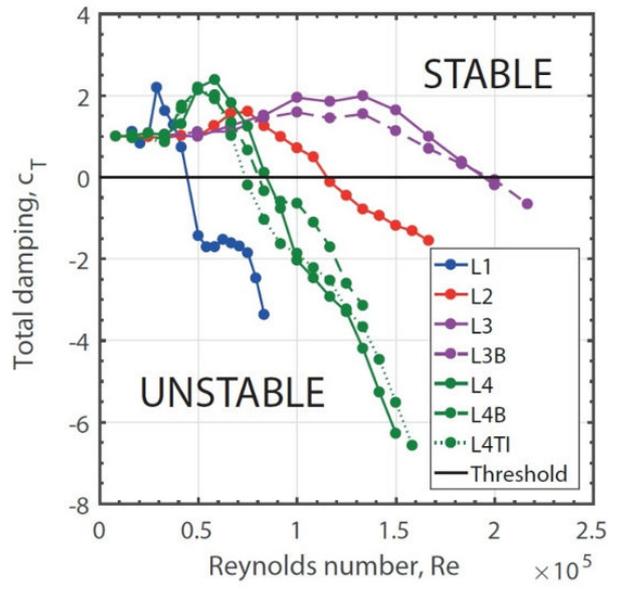

**Figure 8. Evolution of the total damping coefficient $c_T$ with the Reynolds and the Stokes numbers. Indirect method.**

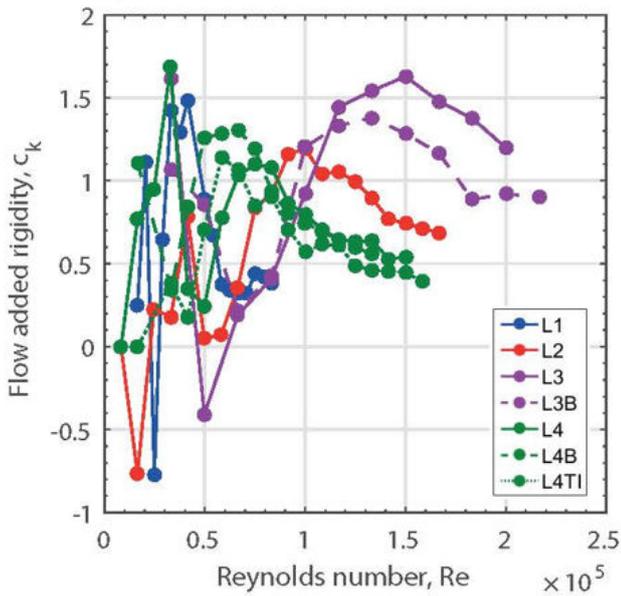

**Figure 7. Evolution of the flow added rigidity coefficient $c_K$ with the Reynolds and the Stokes numbers. Indirect method.**

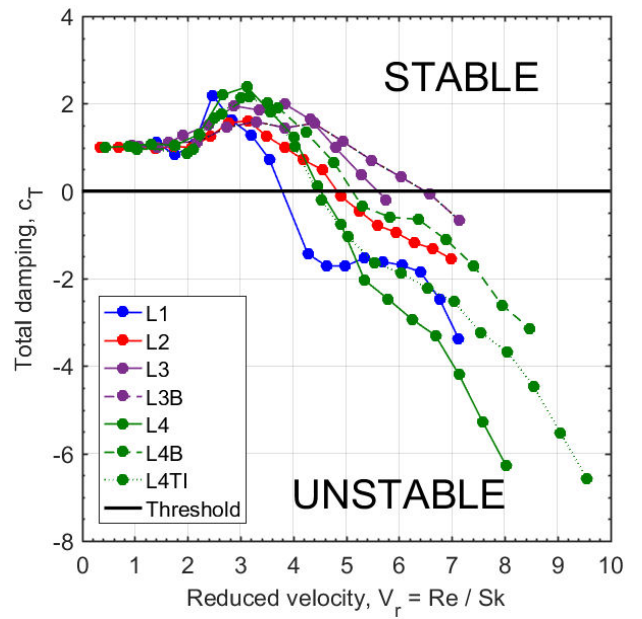

**Figure 9. Evolution of the total damping coefficient $c_T$ with the reduced velocity and the Stokes number. Indirect method**

## CONCLUSIONS

In this work, we have studied the lift vibration of a flexible tube subject to a single phase cross flow. The flexible tube is located in the central position of a square rigid tube bundle. A dimensional analysis shows that the fluid-elastic coefficients are not a function of the reduced velocity only, but also depend on the Reynolds number. This observation is confirmed in our experiments by using two different methods of measurement. In the direct method, an harmonic motion of increasing frequency is imposed to the tube. In the indirect method, the coefficients are obtained from the changes in tube vibration frequency and damping. Both methods suggest the existence of three different dynamics for the flexible tube. At low Reynolds numbers, the fluid velocity has no effect on the stability of the tube. At moderate Reynolds numbers, the tube loses some energy and gets stabilized. At large Reynolds numbers, the tube gains some energy from the fluid and becomes unstable at a critical Reynolds. The experiments show that a tube with a high frequency is more stable than a tube with a low frequency.

Finally, it shall be noted that some significant differences are observed in comparing the results of measurement from the two experimental methods as the Stokes number is increased and as the Reynolds numbers is decreased (i.e. low reduced velocities). We attribute these differences to a bad signal to noise ratio due to some parasitic frequencies in the experimental setup as the forcing frequency of the direct method is increased. Still, these first comparisons are very encouraging and should foster further developments of the direct method.

## ACKNOWLEDGMENTS

The authors acknowledge financial support for this work, which was performed in the framework of a joint research program co-funded by FRAMATOME, EDF and CEA (France).